\documentclass{iaus}
\usepackage{graphicx}

\title[Stellar models: a fundamental ingredient for Population Synthesis tools] 
{Stellar models: \\ firm evidence, open questions and future developments}

\author[Santi Cassisi]   
{Santi Cassisi$^1$}

\affiliation{$^1$ INAF - Astronomical Observatory of Teramo, Via M. Maggini, 64100 Teramo, Italy \\ email: {\tt cassisi@oa-teramo.inaf.it} }

\pubyear{2009}
\volume{262}  
\pagerange{1--10}
\jname{Stellar Populations - Planning for the Next Decade}
\editors{G. Bruzual \& S. Charlot, eds.}
\begin{document}

\maketitle

\begin{abstract} 

During this last decade our knowledge of the evolutionary properties of stars has significantly improved. 
This result has been achieved thanks to our 
improved understanding of the physical behavior of stellar matter in the thermal regimes 
characteristic of the different stellar mass ranges and/or evolutionary stages.

This notwithstanding, the current generation of stellar models is still affected by several, not negligible, 
uncertainties related to our poor knowledge of some thermodynamical processes and  
nuclear reaction rates, as well as the efficiency of mixing processes.
These drawbacks have to be properly taken into account when comparing theory with observations, to derive
evolutionary properties of both resolved and unresolved stellar populations.

In this paper we review 
the major sources of uncertainty along the main evolutionary stages, and emphasize 
their impact on population synthesis techniques.

\end{abstract}

\section{Introduction} 

As far as stellar model input physics is concerned,
significant improvements have been achieved
in the determination of the Equation of State (EOS) of the stellar
matter, opacities, nuclear cross sections, neutrino emission rates. 
At the same time, stellar models computed with this updated physics have been 
extensively tested against the latest observational constraints. 

The capability of the latest generation of stellar models to account for
all the evolutionary phases observed in stellar clusters is undoubtedly 
an exciting achievement, that crowns with success the development of 
stellar evolutionary theories as pursued during the whole second half of the last
century. Following this success, one is often tempted to use
evolutionary results in an uncritical way, i.e., by taking these results at
face value, without accounting for the associated uncertainties. 
However, theoretical uncertainties do exist, as it is clearly shown by the
not negligible differences among the results obtained by different research groups. 

A careful discussion of the uncertainties affecting stellar models for low-mass stars was early addressed by
Chaboyer (1995), who investigated the reliability of  
theoretical models for H-burning stars presently evolving 
in galactic globular clusters (GGCs). This type of investigation 
has been extended to later evolutionary phases by Cassisi et al. 
(1998, 1999), Castellani \& Degl'Innocenti (1999), and Gallart et al. (2005). A discussion of the drawbacks 
of stellar
models for low-mass stars and their impact on widely employed age, distance and chemical composition
diagnostics has been also provided by Cassisi (2005 and references therein), whilst the same 
issues in case of intermediate-mass stellar models have been reviewed by Cassisi (2004).

Stellar evolution models represent also a key ingredient for stellar
population synthesis (SPS) tools, applied to the study of unresolves stellar populations. 
The accuracy and reliability of the adopted evolutionary
framework affects therefore also our ability to derive physical properties of galaxies when 
employing SPS techniques.
Until a few years ago, the stellar libraries adopted in SPS modelling were used with an uncritical approach,  
without taking 
care of their level of accuracy and completeness. Luckily enough, in recent 
times it is becoming increasingly clear that assessing 
the reliability of the adopted stellar models is a pivotal step to evaluate
quantitatively the uncertainty of SPS results.

\section{On the impact of stellar models uncertainties on SPS predictions}

Stellar models can be affected by significant uncertainties that will affect SPS predictions. 
However, so far, stellar models have been usually emploted by the SPS community in an uncritical way. 
As a consequence, when applying SPS models to 
observations
of both resolved and unresolved stellar populations, no one has considered   
the contribution to the systematic error that  
can affect, for instance, the derived ages and metallicities, 
coming from systematic uncertainties in the adopted model library.

The main reasons for this neglect are most probably due to the following: 
i) to check the effects of stellar model uncertainties on SPS predictions is an extremely 
time-consuming procedure - since one has to verify separately 
how the uncertainties in the various evolutionary stages affect the
various SPS indicators (integrated colors, spectra and photometric indices); ii) it is 
sometime extremely difficult to obtain realistic estimates of the \lq{true}\rq\ errors affecting the 
various stellar model 
predictions as evolutionary lifetimes, luminosity and effective temperature; iii) 
to test the impact of different, independent 
stellar model libraries is often problematic due to the difficulty of 
incorporating a stellar library in a SPS code.

This notwithstanding, there is now an ongoing effort in this direction 
(Gallart et al.~2005, Coelho et al.~2007, Lee et al.~2009a, 2009b, 
Cenarro et al.~2008, Conroy et al.~2009, Percival \& Salaris~2009, and references therein). For a detailed 
discussion, we refer the interested reader to the quoted papers. However, we wish to summarize here the 
main conclusions of these analyses. Some of the quoted investigations have shown that the various SPS 
diagnostics are affected differently and, sometimes, in the opposite sense, by systematic changes in the stellar 
model predictions, such as luminosity and/or effective temperature, and slight offsets between the 
metallicity scales of the adopted stellar model set and spectral library. This occurrence has a noteworthy  
implication for methods which fit simultaneously to several spectral indices for deriving ages and
 metallicities of unresolved stellar populations, since a failure to match several indices simultaneously could, 
spuriously, be interpreted for example as an indication of a non scaled-solar heavy elements distribution. 
It has been also proven that the inclusion of the Asymptotic Giant Branch (AGB) stage in SPS models   
is fundamental for understanding the physical properties of galaxies. However, a different treatment of this  
uncertain evolutionary stage alters the final results significantly as, for instance, the inferred galaxy masses  
(Bruzual 2007). 
Therefore, SPS models that do not account for the current uncertainties in AGB modelling are largely 
underestimating errors and may even be introducing systematic biases.

As for the possibility of testing the impact of independent stellar model databases in SPS tools, 
the situation has significantly improved in these last few years thanks to the availability of new, updated and complete sets of stellar models (Pietrinferni et al. ~2004, 2006, Dotter et al.~2007, Bertelli et al.~2008 and references therein) that can be easily incorporated in a SPS code. 
Ideally, the SPS community should now make the effort of considering these 
different stellar model libraries in the SPS codes, 
in order to evaluate the effect of using independent model prescriptions on their SPS results.

The final point that has to be addressed is how to provide a realistic estimate of the uncertainties  
affecting the various stellar models predictions. Clearly, this information has to be provided by people working 
in the field of stellar evolution, by checking the impact of different physical inputs and/or physical 
assumptions on their own model computations, and comparing models provided by various authors. 

\section{Stellar models: the state-of-the-art}
\label{sta}

An in-depth discussion of the accuracy and reliability of contemporary stellar model 
predictions would be clearly beyond the scope of  
this paper. For such a reason, we wish to address only the major open problems affecting the 
model computations. In particular, we want to emphasize the uncertainties associated with the 
treatment of mass loss during the 
\lq{cool}\rq\ evolutionary stages, i.e. the Red Giant Branch (RGB) and the AGB, 
and evolution of stars during the Thermal Pulses stage (TPAGB).  
However, we consider also useful to 
provide some indications about the level 
of reliability of model predictions concerning \lq{less problematic}\rq\ evolutionary stages:

\subsection{The central H-burning stage}

In the last decade, the accuracy of central H-burning (main sequence) theoretical models has improved a lot. 
This occurrence is due to the availability of updated and accurate predictions concerning both 
the thermal and opacitive properties of matter in the relevant regime for both the interiors and 
atmospheres of low- and intermediate-mass stars. 
Some residual uncertainty is associated to (some) nuclear reaction rates. 
A large effort has been devoted to improve the measurements at energies as close as
possible to the Gamow peak, i.e. the energies at which nuclear reactions occur in the
stars. Thanks to this effort the nuclear processes involved in the {\sl p-p} chain have a 
small uncertainty. As a consequence, the uncertainty on the age - luminosity calibration 
is also negligible ($<$2\%).
However, near the end of the main sequence, due to the paucity of H, the energy supplied by the
H-burning becomes insufficient and the star reacts contracting its core in order to produce the
requested energy via gravitation. As a consequence, both central temperature and density
increase and, when the temperature attains a value of $\sim15\times10^6K$, the
H-burning process is controlled by the {\sl CNO} cycle, whose efficiency is
critically dependent on the $^{14}N(p,\gamma)^{15}O$ reaction rate, since
this is the slowest reaction of the whole cycle. 
 
Until few years ago, the rate for the $^{14}N(p,\gamma)^{15}O$ reaction was uncertain, by a factor of 5 at least, 
because all available laboratory measurements were 
performed at energies well above the range of interest for astrophysical purposes (Angulo et al. 1999).
The LUNA experiment (Formicola et al. 2003) has significantly improved
the low energy measurements, obtaining an estimate which is about a factor
of 2 lower than previous determinations. 
This lower rate for the $^{14}N(p,\gamma)^{15}O$ reaction leads to a brighter and hotter Turn Off for a fixed
age. The impact of this new rate on the age - luminosity relation (Imbriani et al. 2004) is the following: for a
fixed TO brightness the new calibration predicts systematically older cluster ages, $\sim$0.9Gyr on average.

\subsection{The Red Giant Branch} 
   
RGB stars are cool, reach high luminosities, their evolutionary 
timescales are relatively long, and therefore provide a major contribution to the integrated   
bolometric magnitude and to integrated colors and spectra at wavelengths larger than about 900 nm in old, 
unresolved stellar populations (e.g. Renzini \& Fusi-Pecci~1988; Worthey~1994).   
A correct theoretical prediction of the RGB spectral properties and colors is thus of paramount importance 
for interpreting observations of distant stellar clusters and galaxies   
using population synthesis methods, but also for determining the ages of resolved stellar systems by 
means of isochrone fitting techniques.   

Both the RGB location and slope in the CMD are strongly sensitive to the metallicity, and for this reason,
they are widely used as metallicity indicators.
 
\begin{figure}[b]
\begin{center}
 \includegraphics[width=2.7in]{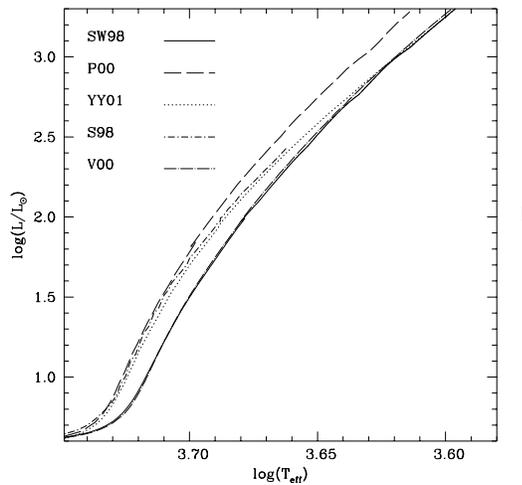} 
 \caption{Comparison between the theoretical RGB tracks provided by various authors 
for the same assumptions about the age and 
chemical composition (see text for more details) (from Salaris et al.~2002).}
   \label{rgbs}
\end{center}
\end{figure}

The $I$-band brightness of the tip of the RGB (TRGB) provides a     
robust standard candle, largely independent of the stellar age and initial   
chemical composition, which allows to estimate distances out to about 
10~Mpc using $HST$ observations. Due to the lingering   
uncertainties on the empirical determination of the TRGB brightness   
zero point, RGB models provide an independent calibration of this important standard candle   
(Salaris \& Cassisi~1997, 1998). 
Moreover, theoretical predictions about the structural properties of RGB stars at the Tip
of the RGB play a fundamental role in determining the main evolutionary properties of their progeny: the
core He-burning stars during the Horizontal Branch (HB) evolutionary phase. In particular,     
HB luminosities (like the TRGB ones) are mostly determined by the value of the electron degenerate    
He-core mass ($M_{core}^{He}$) at the end of the RGB evolution.   
   
Predicted evolutionary timescales along the RGB phase play also a    
fundamental role in the determination of the initial He abundance 
of globular cluster stars through the R parameter    
(Salaris et al.~2004 and references therein).    
  
A detailed analysis of the existing uncertainties in theoretical RGB models, and of the level of
confidence in their predictions has been performed by Salaris et al.~(2002).
As far as the location and slope of RGB evolutionary tracks is concerned, 
model predictions are affected by: the EOS, the low-T opacity, the efficiency of superadiabatic convection, 
the choice about the model outer boundary conditions and the initial chemical
abundances. It has been already emphasized that in the thermal regime appropriate for RGB stars, big 
improvements have been achieved concerning both the EOS and low-T opacity. 

We think that it is worthwhile to discuss more in detail the issue related to the efficiency of the outer 
convection. As well known, the efficiency of superadiabatic convection parametrized by the 
mixing length parameter ($\alpha_{\rm MLT}$)
is usually calibrated by reproducing the solar $T_{\rm eff}$, and this   
solar-calibrated value is then used for stellar models of different masses and along different evolutionary phases,
including the RGB one. 
The adopted procedure guarantees that models always predict correctly the $T_{\rm eff}$  
of at least solar type stars. However, 
the RGB location is much more sensitive to the value of $\alpha_{\rm MLT}$   
than the main sequence. Therefore, it is important to verify that a solar $\alpha_{\rm MLT}$ is  
suitable also for RGB stars of various metallicities.   

A source of concern about an {\sl a priori} assumption of a solar $\alpha_{\rm MLT}$   
for RGB computations comes from the fact that recent models from various authors, all using a suitably     
calibrated solar value of $\alpha_{\rm MLT}$, do not show the same RGB temperatures.    
This means that -- for a fixed empirical RGB temperature scale --   
the calibration of $\alpha_{\rm MLT}$ based on RGB $T_{\rm eff}$ estimates values   
would not provide always the solar value.   
Figure~\ref{rgbs} displays several isochrones produced by different  
groups, all computed with the same initial chemical composition, same opacities,    
and the appropriate solar calibrated values of $\alpha_{\rm MLT}$: Vandenberg et al. (2000, V00) and Salaris
\& Weiss (1998, SW98) models are identical, the Padua ones (Girardi et al. 2000, P00) are systematically hotter 
by $\sim$200 K, while the YY01 ones (Yi et al.~2001) have a different shape.  
This comparison shows clearly that if one set of solar calibrated RGBs  
can reproduce a set of empirical RGB temperatures, the others cannot,  
and therefore in some case a solar calibrated $\alpha_{\rm MLT}$  
value may not be adequate.  
The reason for these discrepancies must be due to some    
difference in the input physics, like the EOS and/or the boundary conditions,    
which is not compensated by the solar re-calibration of $\alpha_{\rm MLT}$
(see also Vandenberg et al.~2008).

This occurrence clearly points out the fact that one  
cannot expect the same RGB $T_{\rm eff}$ from solar calibrated models  
that do not employ exactly the same input physics. Therefore  it is always necessary to compare 
RGB models with observations  
to ensure the proper calibration of $\alpha_{\rm MLT}$ for RGB stars.  In the meantime, 
from the previous comparison, we can safely estimate that current uncertainties 
on the $T_{eff}$ scale of RGB models are of the order of $200-300$~K.

An other important prediction provided by RGB models is the number of stars in any given bin of the RGB 
luminosity function (LF -- star counts as a function of brightness) which is determined by the 
local evolutionary rate. Therefore, the comparison
between empirical and theoretical RGB LF represents a key test for the accuracy of the predicted RGB 
timescales (Renzini \& Fusi Pecci 1988). In addition, there are many more reasons why an investigation 
of RGB star counts is important, for instance: i) being the RGB stars among the brightest objects in a galaxy, 
their number has a strong influence on the integrated properties of the galactic stellar population; ii) 
the number ratio between RGB and
stars along the AGB can be used to constrain the Star Formation History of a galaxy (Greggio
2000).

The investigation of the accuracy of theoretical RGB LFs has been performed by Zoccali \& Piotto (2000)
by adopting a large database of GGC RGB LFs. The main outcome of their analysis was the evidence of, on average,
a good agreement, in the whole explored metallicity range, between observations and the adopted theoretical 
predictions (but see also Sandquist \& Martel~2007). 

\begin{figure}[b]
\begin{center}
 \includegraphics[width=3.3in]{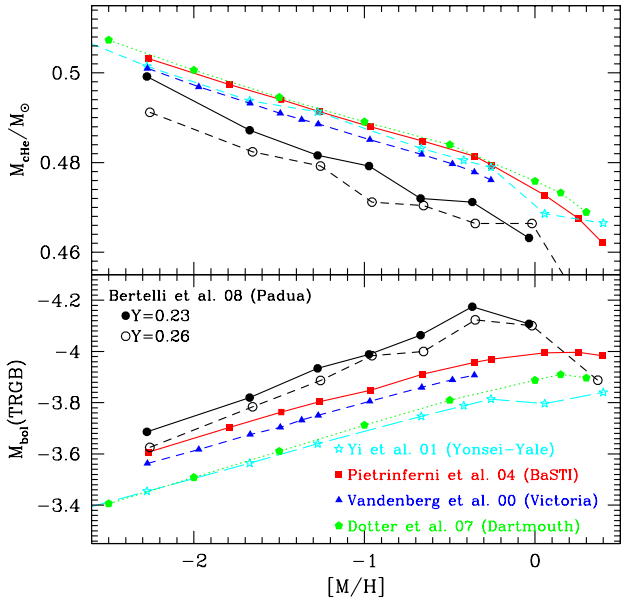} 
 \caption{The trends of $M_{core}^{He}$ and $\rm M_{bol}^{TRGB}$ as a function of the metallicity as provided by 
the most recent stellar model libraries (see text for more details). }
   \label{trgb}
\end{center}
\end{figure}

An important quantity provided by models is the luminosity  of the TRGB.
The observational and evolutionary properties of RGB stars at the TRGB play a pivotal role in current 
stellar astrophysical
research. As mentioned before, the reasons are manifold: i) the mass size of the He core at the He flash 
fixes not only the TRGB brightness but
also the luminosity of the Horizontal Branch, ii) the TRGB brightness is one of the 
most important primary distance indicators.

As for the uncertainties affecting theoretical predictions about the TRGB brightness, it is clear that, being this quantity
fixed by the He core mass, any uncertainty affecting the predictions of 
$M_{core}^{He}$ immediately translates into an error on $\rm M_{bol}^{TRGB}$.
An exhaustive analysis of the physical parameters that affect the estimate of $M_{core}^{He}$ can be found 
in Salaris et al. (2002). Suffice to remember that
the physical inputs that have the largest impact in the estimate of $M_{core}^{He}$ are the efficiency of 
atomic diffusion and the
conductive opacity. Unfortunately, no updates are available concerning a more realistic estimate of 
the real efficiency of diffusion in low-mass stars, apart from the Sun.
On the contrary, concerning the conductive opacity, large improvements have been obtained recently 
(Potekhin 1999, Cassisi et al. 2007). This new set represents a significant improvement 
(both in the accuracy and in the range of validity) with respect to previous estimates.

We show in fig.~\ref{trgb} the comparison of the most recent results 
(Bertelli et al. 2008 - Padua, Pietrinferni et al. 2004 - BaSTI, Vandenberg et al. 2000 - Victoria, Dotter et al. 2007 - 
Dartmouth, Yi et al. 2001 - Yonsei-Yale) concerning     
the TRGB bolometric magnitude and $M_{core}^{He}$ at the He-flash; the   
displayed quantities refer to a 0.8$M_{\odot}$ model and various   
initial metallicities.   
When excluding the Padua models, there exists a fair agreement among the various predictions about 
$M_{core}^{He}$: at fixed metallicity the 
spread among the various sets of models is at the level of $0.003M_\odot$. For the Padua models, 
we show the results corresponding to the two different initial He contents adopted by the authors: we have no clear explanation 
for the fact that the Padua models predict 
the lowest values for $M_{core}^{He}$, as well as for the presence of an \lq{erratic}\rq\ behavior of the values corresponding to the 
different He abundances: for a fixed total mass and metallicity, the $M_{core}^{He}$ value is expected to be a 
monotonic function of the initial He abundance. Concerning the trend of  $\rm M_{bol}^{\rm TRGB}$, all model 
predictions at a given metallicity are in agreement within $\sim 0.15$ mag, with the exception    
of the Padua models that appear to be brighter, at odds with the fact that they predict the lowest $M_{core}^{He}$  values.
In case of the Yonsei-Yale models, the result is also surprising since the fainter TRGB luminosity cannot be explained
by much smaller $M_{core}^{He}$ values, because this quantity is very similar to, for instance, the results given by Vandenberg et
al.~(2000).
When neglecting the Padua and Yonsei-Yale models, the $\sim0.1$ mag   
spread among the different TRGB brightness estimates can be interpreted in terms of differences in the adopted physical inputs
such as for instance the electron conduction opacities.

\begin{figure}[b]
\begin{center}
 \includegraphics[width=2.7in]{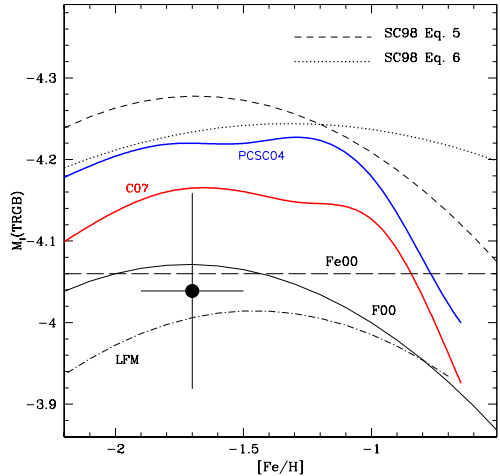} 
 \caption{A comparison among theoretical calibrations of the I-Cousins magnitude of the TRGB as provided by
Pietrinferni et al.~(2004, PCSC04) and Salaris \& Cassisi~(1998, SC98, see their equations~5 and 6), 
and (semi-)empirical ones as given by Lee et al.~(1993, LFM), Ferrarese et al.~(2000, Fe00) and Ferraro et al.~(2000, F00). 
The new theoretical calibration by Cassisi et al. (2007, C07) is also shown. The full circle with the error bars corresponds to the empirical 
calibration provided by Bellazzini et al.~(2001).}
   \label{itrgb}
\end{center}
\end{figure}

Due to its relevance as standard candle, it is worthwhile showing a comparison between theoretical
predictions about the I-Cousins magnitude of the TRGB and empirical calibrations. This comparison is displayed in
fig.~\ref{itrgb}, where we show also the recent empirical calibration provided by Bellazzini et al. (2001) based on the well 
populated GGC $\omega$~Cen. In this plot, we have shown different calibrations of $M_I^{TRGB}$ as a function of the metallicity based on our own
stellar models. These calibration are about $0.20$ mag brighter than the most recent, empirical ones. However, it is also important to note the new calibration 
by Cassisi et al.~(2007) - based on the new conductive opacity - is in better agreement with the empirical evidence.

\section{Stellar models: the open problems}

\subsection{The mass-loss efficiency}

One of the thorniest problems in current stellar evolution theory is that related to the efficiency of mass loss (ML) during both the RGB and AGB stage. 
In fact, the efficiency of ML during the RGB strongly controls the 
$T_{eff}$ - and hence the color - of the star along the HB stage, while during the AGB by reducing the envelope mass, it truncates the AGB evolution  - 
and hence the contribution of the star to the infrared flux of the global stellar population.

The astrophysical impact of ML in both Pop. I and II giants is huge and affects also the interpretation of 
the UV excess in ellipticals, or the interaction between the cool intracluster medium and hot halo gas. 
There is so much indirect but quantitative evidence for ML  during the giant branches evolution, namely the HB morphology 
and the 2$^{nd}$ parameter problem, the pulsational properties of RR Lyrae, 
the absence of AGB stars brighter than the RGB tip and the masses of White Dwarfs in GCs (see Catelan 2009 and references therein).
However, despite its importance, complete empirical determinations  as well as a comprehensive physical description of the involved 
processes are still lacking. So far, there is a lack of any empirical law directly calibrated on Population II 
giants.  Indeed, only a few, sparse estimates of ML for giants along the brightest portion of the RGB and AGB exist. From a theoretical point of view, 
our knowledge of the ML timescales, driving mechanisms, dependence on 
stellar parameters and metallicity is also very poor. The consequence is that there is little theoretical 
or observational guidance on how to incorporate ML into stellar model computations.

Without a better recipe, models of stellar evolution  incorporate ML by using analytical formulae calibrated on Population I 
bright giants, the first and most used being the Reimers (1975) formula, extrapolated towards lower luminosity by also 
introducing a free parameter $\eta$ (typically 0.3), to account for a somewhat less efficient ML along the RGB. 
A few other formulae, which are variants of the Reimers one, have been proposed in the subsequent years 
(see Catelan 2009 for a detailed discussion on this issue) but there is no {\sl a priori} reason for choosing among the different alternatives. 
In these last few years, at least on the observational side, the situation is improving thanks to the availability of the observational facilities 
associated to, for instance, the SPITZER telescope. As a consequence, there is growing amount of empirical data concerning 
ML estimates for Pop. II red giants (see Origlia et al. 2007 and references therein). 

The preliminary empirical ML law that has been obtained (Origlia et al. 2007) appears significantly different 
(flatter) than the Reimers formula, that appears to be ruled out by current empirical estimates at 
the $3\sigma$ level. In addition, it seems that the ML phenomenon is not a continuous process along the RGB but an episodic phenomenon, and it does 
not appear to be strongly correlated with the cluster metallicity.

The situation is still more controversial and complicated in the case of AGB stars, due to the link 
existing between the ML efficiency (and the physical processes that causes the ML) and the evolutionary, 
structural and pulsational 
properties of the evolving star (see van Loon 2008 for a detailed review on this issue). 

We expect that, thanks to the availability of new observational facilities and the huge theoretical 
and observational effort that is devoted to this subject, 
significant improvements could be obtained in the next decade. This occurrence would contribute, 
not only to obtain more accurate stellar models, but also in reducing the degrees of freedom in the SPS modeling. 

\subsection{The AGB stage modeling}

The computation of AGB stellar models is one of the most complicated task for the stellar evolution community. The reasons for this has to 
be found in the realization that the evolutionary properties of these stars are hugely dependent on the complex link existing among mixing processes (the 3DU),
ML efficiency, nucleosynthesis and envelope opacity stratification. The evolutionary results that can be obtained, 
strongly depend on the assumptions about the efficiency of these various processes - and their treatment in the numerical codes - 
adopted in the model computations.

Concerning the 3DU, in spite of its fundamental relevance in determining the chemical enrichment of TPAGB star envelopes, 
its treatment in stellar evolutionary code is quite uncertain. This is due to the fact that we are not yet able to properly 
describe convection inside the stars, and in particular, in case of mixing occurring in a region with a strong  
composition/opacity discontinuity. Many different - arbitrary - methods can be envisaged to treat the occurrence 
of the 3DU, but each one of these approaches has no robust physical ground. This has the important implication
that in all fully evolutionary AGB models, the efficiency of the 3DU is managed by using one  
(or more) free parameter(s).

An important physical implication of the occurrence of the 3DU is the huge change in the envelope C/O ratio that 
is caused by this process (Ventura \& Marigo~2009). The change in the C/O ratio when it approaches 
(and overcomes) unity has huge effects on the opacitive properties of the stellar envelope (Marigo~2002). 
As a consequence, the C/O ratio drives sharp discontinuities in many observational properties of AGB stars: 
$T_{eff}$, colors, spectra, mass loss efficiency, etc. After many years during which only approximate evaluations 
for the C-rich mixture opacity were available, the situation is now largely improving and new opacity tables for AGB 
envelopes can be now incorporated in evolutionary codes (Cristallo et al. 2008, Weiss \& Ferguson 2009). 
These new opacities, although in good qualitative agreement with the previous estimates, show significant, 
quantitative differences. This occurrence has the effect that the $T_{eff}$ scale for AGB models is expected 
to be significantly affected, with consequences on SPS modeling that are still to be fully exploited.

Before concluding, we wish to comment briefly on synthetic AGB modeling (see, e.g., Marigo et al. 2008). 
These synthetic simulations are based on analytical, functional relations providing the link among the main structural - such as the He core mass - 
evolutionary properties - as $T_{eff}$ and luminosity - and chemical properties - mainly the metallicity and the C/O ratio -, obtained from the
 numerical integration of the stellar structure
and envelope. The (recent) literature - see also the proceedings of this meeting - contains several claims about the fact that the 
new synthetic AGB computations are significantly better and more accurate than previous ones, allowing a better (or \lq{perfect}\rq!) match to various 
empirical constraints. 
In our belief, these claims are somewhat misleading. Even though large improvements have been achieved in the field of AGB star modelling (see above), 
the same problems affecting the full evolutionary models affect also synthetic AGB modeling. 
As a consequence, in these synthetic simulations there is a number of free parameters that have to be tuned 
\lq{by hand}\rq\ in order to reproduce some observational 
constraints such as for instance the Magellanic Clouds C-stars luminosity functions. Synthetic AGB models 
constitute a much faster way to produce approximate models with 
these parameters tuned to match some observations, but it is not clear what their predictive power is, outside 
the parameter space covered by the calibrating AGB populations.
We think this is an important warning that should be kept in mind by the SPS 
community, in order to avoid 
overestimating the reliability of the SPS tools that incorporate the new generations of synthetic AGB models.

It will not be possible to have a realistic and physically well grounded treatment of the AGB stage, until we 
will account for the various physical 
processes at work in AGB stars starting from a robust theory, free of tunable parameters.

\acknowledgements           
We warmly acknowledge M. Salaris for his pertinent suggestions and for reading an
early draft of this manuscript. We are also grateful to him for many enlightening discussions. We thank also A. Pietrinferni for his continuous help. We warmly thank  the LOC and the SOC for organizing this stimulating meeting.


\end{document}